# Fabrication and performance tests of a segmented p-type HPGe detector

George S. King III[a], Frank T. Avignone III[a], Christopher E. Cox[b],
Todd W. Hossbach[a,], Wayne Jennings[b], and James .H. Reeves[c]

[a] Department of physics and Astronomy, University of South Carolina, Columbia, South Carolina 29208, USA
[b] Princeton Gamma-Tech Instruments, Inc. Princeton, New Jersey 08540, USA
[c] Reeves and Sons, LLC. Richland, Washington 99354, USA

**Abstract**

A p-type HPGe detector has been segmented by cutting, with a diamond saw, and etching six circumferential grooves through the Li-diffused dead layer. The degree of segmentation was tested with the 88 keV gamma rays from a well-collimated source of $^{109}$Cd. The fraction of events, recognized as occurring in more than one segment, and rejected from the energy interval 2038±5 keV, was measured as 0.59.

## 1. Introduction

High-purity germanium (HPGe) gamma-ray detectors are key tools in many areas of nuclear physics and associated technologies. In areas of fundamental physics they have found application in: $\beta\beta$–decay [1,2], searches for Cold Dark Matter [3-6], and searches for axions [7-9]. In these specific applications the expected events would create electron-hole clusters at a single site in the crystal. Much of the background is due to gamma rays that undergo multiple scattering, thereby creating several electron-hole clusters at multiple sites. This common source of background can be identified by pulse-shape analysis [10-14]. The clusters rapidly reach their terminal velocities; hence the drift times to their respective terminals vary. This effect creates characteristic time dependences in the displacement currents that allow one to distinguish a significant fraction of multiple-site events from single-site events. This is commonly referred to as "pulse-shape discrimination" (PSD).

The same thing can be accomplished by separating the detector into a number of electrically-isolated segments, each responding independently to events that occur within it. Events that create electron-hole clusters in more than one segment (multi-site events) are then easily identified. Since both of these techniques are independent, they can be applied simultaneously. Application of PSD, following segmentation cuts, for example, could be used to identify events in which multiple cluster sites are created within the same segment. The combination of these two techniques has been proposed as tool for a background reduction by both the GERDA [15], and Majorana [16], $^{76}Ge$, $0\nu\beta\beta$–decay collaborations. In this article we discuss a specific technique for segmenting p-donor, HPGe detectors.

**2. Detector segmentation**

All of the commercially available segmented HPGe detectors are fabricated from n-donor germanium. The outer surface is masked and ion-implanted with boron to form the electrical contacts (segments). The degree and geometry of the segmentation is dictated by the mask. N-type detectors are more difficult to grow and as a result are more costly. This is especially true of the larger detectors, for example 80 × 80 mm semi-coaxial detectors with about 100% of the efficiency of a 76.2 × 76.2 mm NaI(Tl) detector for 1333 keV gamma rays. Detectors made from p-donor germanium are more easily grown, therefore less expensive. Semi-coaxial p-type detectors have a Li-diffused "dead layer" over a great deal of the outer surface. In some configurations, this includes the closed end, the cylindrical surface and the back surface, up to a narrow ditch of high-resistance surface separating the conducting n-type Li-diffused outer surface from the conducting metal implanted inner surface of the central hole. P-donor detectors have been segmented in the past by masking the outer surfaces prior to the Li-diffusion process [17]. This has also been successful in segmenting Ge planar detectors [18]. The disadvantage of this technique is that the lithium will migrate if the detector remains warm for any length of time. In addition the degree of segmentation may depend on the details of the Li-diffusion process. P-type HPGe detectors have also been segmented with amorphous Ge contact barriers [19].

A new technique was developed whereby a closed end, semi-coaxial p-type detector, with a well measured Li diffused dead layer, was segmented by cutting circumferential grooves through the dead layer followed by a chemical etch of the grooves.

The Li-diffused layer was ~0.5 mm deep. The grooves were ~0.5 mm wide and ~3 mm deep. They were cut with a 0.41 mm thick diamond saw, mounted on a milling machine. The crystal was mounted and centered on a rotating table. The groves were cut by making several shallow cuts around the circumference until a total depth of 3.0 mm was achieved. Ice was used to cool the crystal and saw during the cutting. The configuration of the detector (called PSEG) is shown in figure 1.

Following preliminary testing, the detector was installed in a specially made cryostat, designed to minimize crosstalk between segments. A general plan of the cryostat is shown in figure 2. The individual contacts, leads and preamplifiers are located around the circumference to minimize crosstalk.

In this case, negative high voltage is applied to the center contact, and the center-contact signal is coupled to the $1^{st}$ stage preamplifier input through a high-voltage capacitor. The signal from each of the five segments is DC coupled to its preamplifier input. With this method of biasing the detector, only one high-voltage capacitor is needed inside the cryostat.

The electronic system, used to operate the detector, utilizes two synchronized DGF4C modules made by X-Ray Instruments Associates (XIA), LLC. The DGF4C has four independent, 14-bit, 40 MHz analog-to-digital converters (ADCs). The ADCs are followed by First-in-First-out (FIFO) buffers capable of storing 1024 ADC values for a single event. A programmable digital filter and trigger logic is in parallel with each FIFO. The DGF4C is a "smart filter" for incoming pulses.

Following the usual tests for depletion voltage, operating voltages, and leakage currents, a well collimated source of 88 keV gamma rays from the decay of $^{109}Cd$ was used to determine the degree of segmentation. A 0.75 mCi source was located in a lead block with a 46.5 mm long and 3 mm diameter collimator hole. The collimated beam of gamma rays was directed at the center of each segment and spectra from all five segments and the center contact were simultaneously collected.

Sample spectra showing the illuminated segment and its neighbors are shown in figures 4 through 8. It is clear from these sample spectra that the detector is well segmented. It should be noted at this point that the centers of the segments were determined by moving the collimated gamma ray beam in steps as the data were recorded.

It is clear from these figures that segmentation of a p-type detector in this manner is very successful. The next tests were to determine if the grooves would perturb the electric field enough to have a significant detrimental effect on pulse shapes. This is important if it is intended to apply both segmentation and PSD cuts. In figures 9 through 14 pulses from single site and from multiple site events are depicted.

From these examples it is clear that this detector could, with the proper electronic system, have the required characteristics for pulse-shape analysis.

The grey and blue portion of the data, from which figure 15 was made, was collected by using a distributed source of $^{232}Th$, and also includes the room background. The red portion of the histogram is what remains after events occurring in two or more segments are removed. No pulse shape analysis was applied. Prior to the segmentation cut there were 173 events in the energy interval, 2038±5 keV. There were 71 events remaining in the interval after the cuts were made. In other words, 59% of the events interacted in multi-segments, and accordingly were background events clearly recognizable as not having the single-site character.

This concludes phase-I of our segmentation R&D. It is clear from the foregoing discussion that p-type Ge detectors can be successfully segmented by this method, and that even this simple configuration is fairly effective in discriminating multi-site from single site events. In addition, it is clear that some level of pulse-shape discrimination will be possible. In the next phase of the program, the front-end electronic system will be optimized for pulse-shape analysis.

A detector segmented in this way should remain cold. If it would remain warm for long periods, the lithium will eventually drift, and could destroy the segmentation; the lithium could begin to join the segments. It is not known at this point how robust the segmentation is against that eventuality. In the last phase of our R&D program, the detector will be cycled several times, to answer this question. Finally, it will be warmed, pumped, and baked as HPGe detectors are frequently treated when the vacuum deteriorates and the leakage current increases.

## 3. Summary and Conclusions

A closed end, p-type semi-coaxial HPGe detector has been segmented by cutting grooves through the Li-diffused dead layer and etching the grooves. The degree of segmentation has been tested, and the efficacy of reducing the background in the interval 2038±5 keV, from the continuum dominated by that from the 2615 keV gamma ray in $^{208}Tl$, has been measured. Segmentation cuts eliminated 59% of these events.

Preliminary data imply that the shapes of the current pulses have not been affected by the segmentation. In the next R&D phase, an attempt will be made to improve the front-end electronic system to optimize the effectiveness of pulse shape analysis. In the final phase, tests will be made of the robustness of the segmentation against thermal recycling as well as heating and pumping.

**Acknowledgements**

This work was supported by the NSF Grants:PHY-0139294, PHY0500337.and by the University of South Carolina Education Foundation. The authors wish to express their appreciation to Drs. Harry Miley, and Craig Aalseth for their interest and advice.

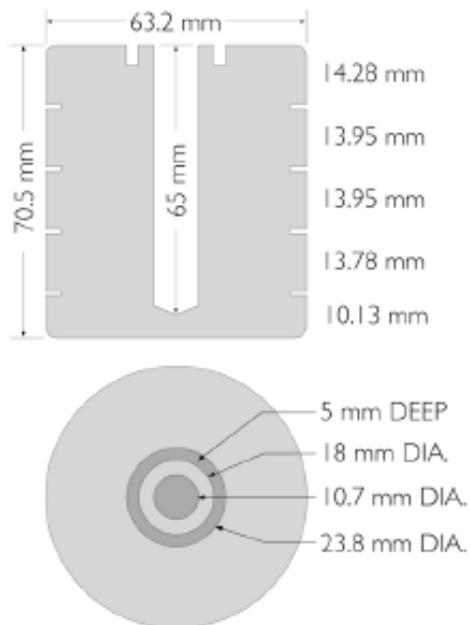

Figure 1.

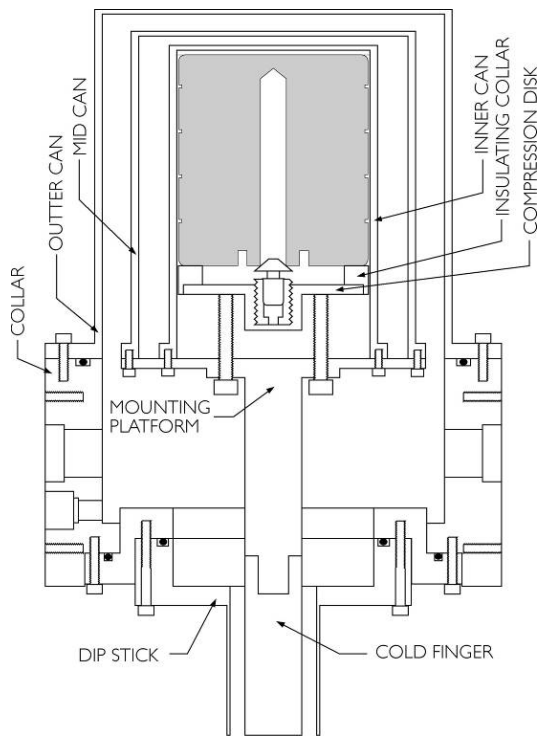

Figure 2.

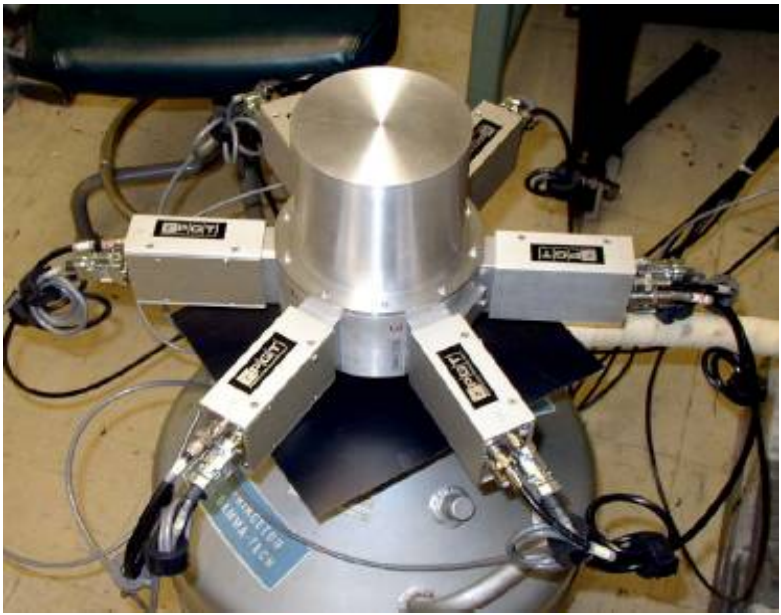

Figure 3.

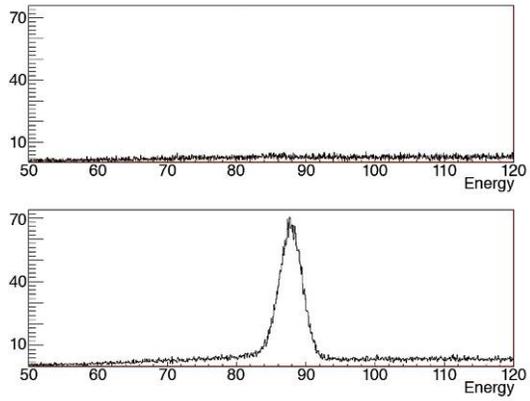

Figure 4.

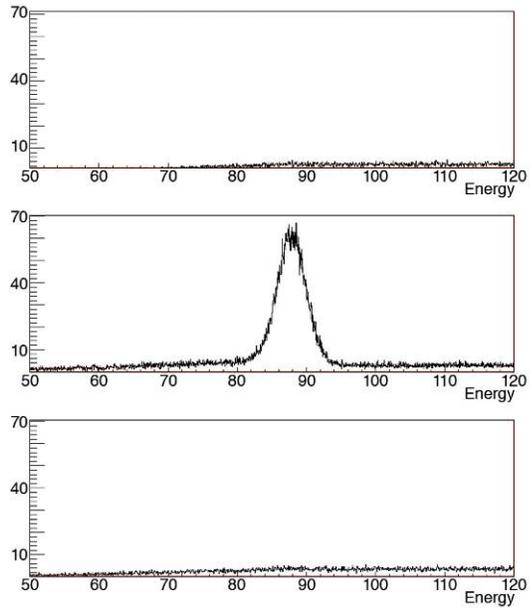

Figure 5.

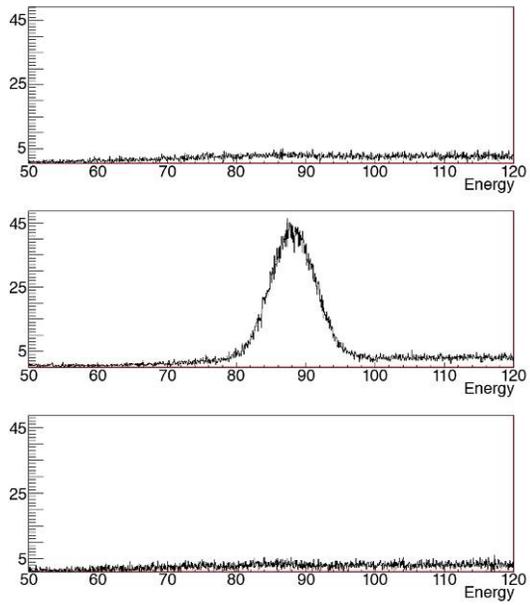

Figure 6.

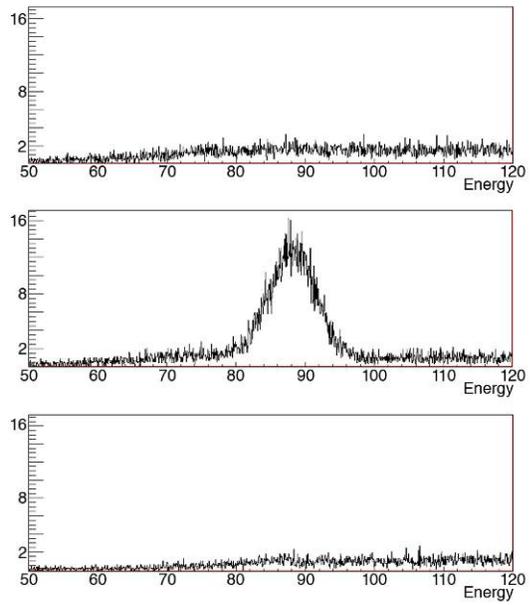

Figure 7.

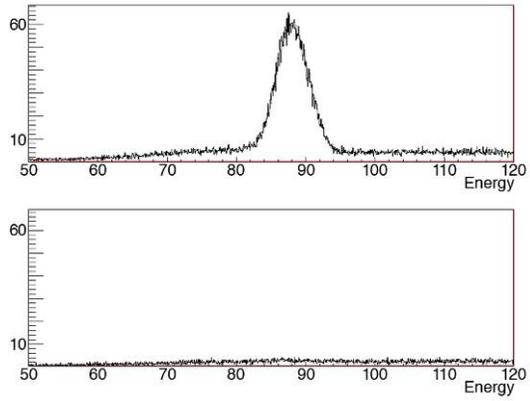

Figure 8.

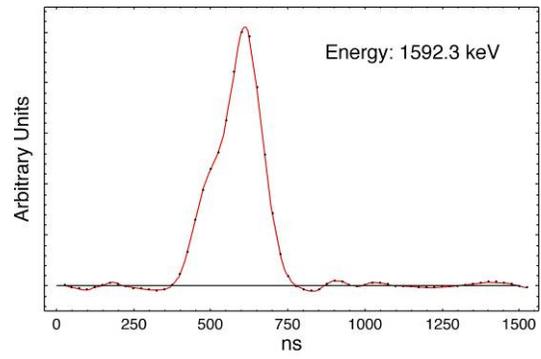

Figure 9.

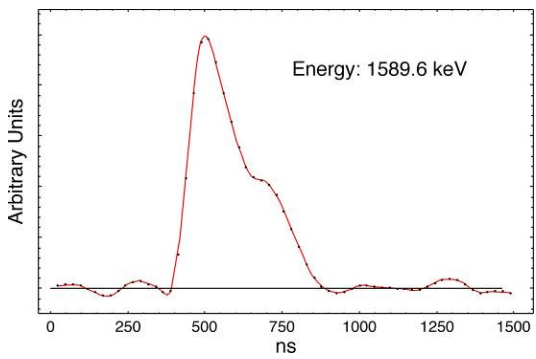

Figure 10.

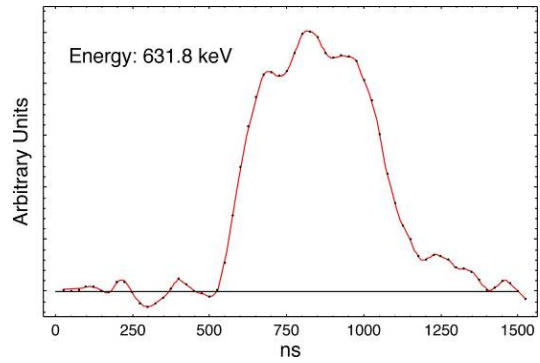

Figure 11.

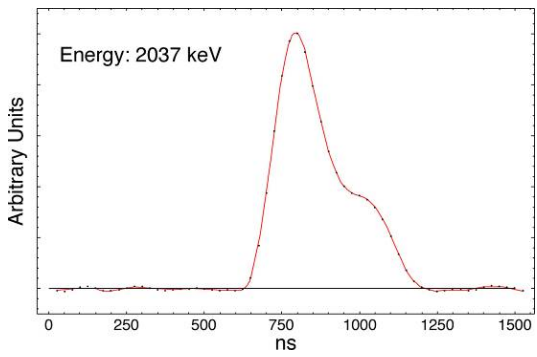

Figure 12.

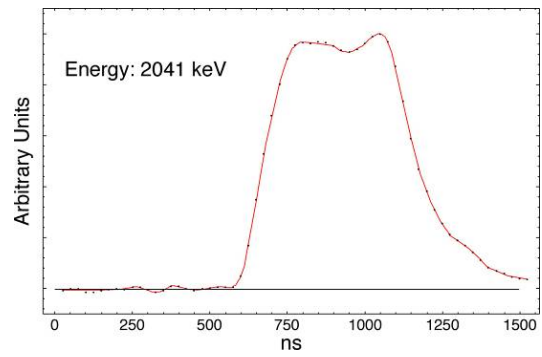

Figure 13.

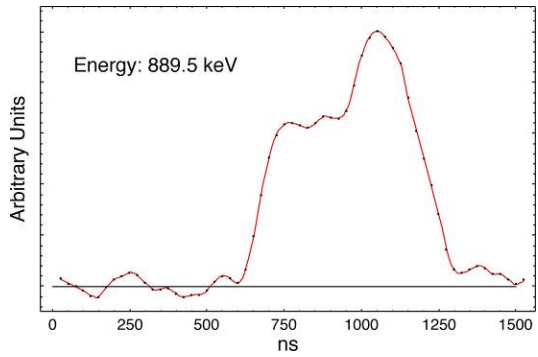

Figure 14.

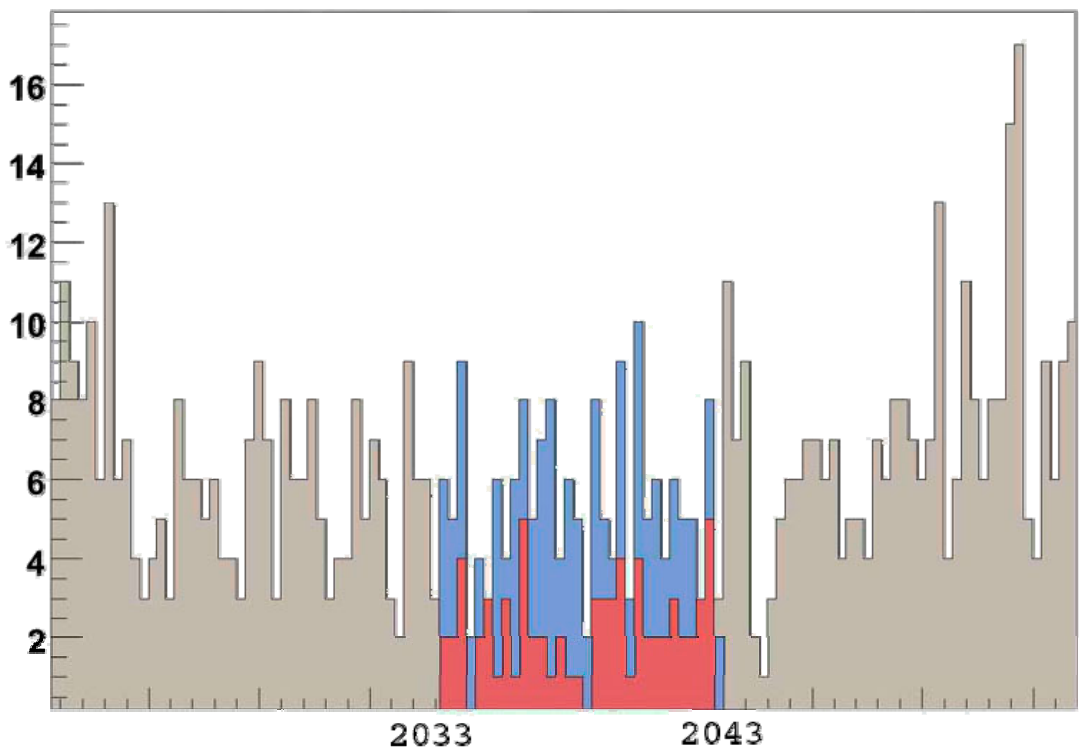

Figure 15.

**Figure Captions:**

Figure 1. The approximate dimensions and configuration of the PSEG detector

Figure 2. An engineering drawing of the PSEG cryostat without dimensions

Figure 3. A photograph of P-Seg in its cryostat and ready for operation.

Figure 4. Spectrum from segment 1 (lower) and segment 2 (upper) when the beam is on segment 1.

Figure 5. Spectra from segments 1,2,and 3, when the gamma ray was directed at segment 2 (middle).

Figure 6. Spectra from segments 2, 3, and 4 when the gamma ray was directed at segment 3 (middle).

Figure 7. Spectra from segments 3,4, and 5, when the gamma ray was directed at segment 4 (middle).

Figure 8. Spectra from segments 4 and 5, when the gamma ray was directed at segment 5 (top).

Figure 9. A differentiated current pulse from the double-escape peak at 1592 keV from the 2615 keV $^{208}Tl$ gamma ray. This must be a single site event. The interaction occurred closer to the outer surface of the cylinder.

Figure 10. Another single-site event pulse from the double escape peak at 1592 keV. This event occurred closer to the hole at axis of the detector.

Figure 11. An example of a pulse from a multi-site event. The energy deposited in the detector was 632 keV.

Figure 12. Another example of a single site event occurring closer to the hole on the axis of the detector. The total energy deposited in this case was 2037 keV

Figure 13. Another example of a pulse from a multi-site event. The total energy deposited was 2040 keV, interestingly close energy deposit expected from the $0\nu\beta\beta$–decay of $^{76}Ge$.

Figure 14. Another example of a pulse from a multi-site event. In this case the deposited energy was 890 keV.

Figure 15. Histogram of data taken with a distributed source of $^{232}Th$. The signal is from the center contact. The energy range of the blue and red areas is 2038±5 keV, the region of interest for experiments searching for $0\nu\beta\beta$–decay of $^{76}Ge$.